\documentclass[twocolumn]{aastex701}

\begin{document}

\title{Optical Spectroscopy of the Most Compact Accreting Binary Harboring a Magnetic White Dwarf and a Hydrogen-rich Donor}

\shorttitle{\textsc{Gaia19bxc}}
\shortauthors{ Galiullin and Rodriguez et al.}

\correspondingauthor{Ilkham Galiullin}
\email{IlhIGaliullin@kpfu.ru}

\author{Ilkham Galiullin}
\affiliation{Kazan Federal University, Kremlevskaya Str.18, 420008, Kazan, Russia}
\email[]{}  

\author{Antonio C. Rodriguez} 
\affiliation{Department of Astronomy, California Institute of Technology, 1200 E California Blvd, Pasadena, CA, 91125, USA}
\email{}  

\author{Kareem El-Badry} 
\affiliation{Department of Astronomy, California Institute of Technology, 1200 E California Blvd, Pasadena, CA, 91125, USA}
\email{}  

\author[]{Ilaria Caiazzo}
\affiliation{Institute of Science and Technology Austria, Am Campus 1, 3400, Klosterneuburg, Austria}
\affiliation{Department of Astronomy, California Institute of Technology, 1200 E California Blvd, Pasadena, CA, 91125, USA}
\email{}

\author[]{Paula Szkody}
\affiliation{Department of Astronomy, University of Washington, 3910 15th Avenue NE, Seattle, WA 98195, USA}
\email{}

\author[]{Pranav Nagarajan}
\affiliation{Department of Astronomy, California Institute of Technology, 1200 E California Blvd, Pasadena, CA, 91125, USA}
\email{}

\author[]{Samuel Whitebook}
\affiliation{Department of Astronomy, California Institute of Technology, 1200 E California Blvd, Pasadena, CA, 91125, USA}
\email{}

\begin{abstract}

Accreting white dwarfs in close binary systems, commonly known as cataclysmic variables (CVs), with orbital periods below the canonical period minimum ($\approx$ 80 minutes) are rare. Such short periods can only be reached if the donor star in the CV is either significantly evolved before initiating mass transfer to the white dwarf (WD) or metal-poor. We present optical photometry and spectroscopy of Gaia19bxc, a high-amplitude variable identified as a polar CV with an exceptionally short orbital period of 64.42 minutes --- well below the canonical CV period minimum. High-speed photometry confirms persistent double-peaked variability consistent with cyclotron beaming, thus indicating the presence of a magnetic WD. Phase-resolved Keck/LRIS spectroscopy reveals strong hydrogen and helium emission lines but no donor features, indicating the accretor is a magnetic WD and the donor is hydrogen-rich, but cold and faint. The absence of a detectable donor and the low inferred temperature ($\lesssim$ 3500 K) disfavor an evolved donor scenario. Instead, the short period and the system's halo-like kinematics suggest Gaia19bxc may be the first known metal-poor polar. Because metal-poor donors are more compact than solar-metallicity donors of the same mass, they can reach shorter minimum periods. Gaia19bxc is one of only a handful of known metal-poor CVs below the canonical period minimum and has the shortest period of any such magnetic system discovered to date.

\end{abstract}

\section{Introduction}

Binary star systems harboring a compact object are important astrophysical laboratories, serving as probes of mass transfer in magnetic and non-magnetic environments, binary effects on stellar evolution, and in the most extreme systems, supernovae and gravitational waves.  The most abundant and nearby systems, due to the initial mass function, are binaries containing a white dwarf (WD). Cataclysmic variables (CVs) are close binary systems consisting of a WD and a Roche lobe-filling main-sequence companion. In such binaries, a WD accretes matter from a donor star via the inner Lagrangian point. CVs can be magnetic systems (polars and intermediate polars (IPs)) or non-magnetic systems (novae, dwarf novae and nova-like variables), depending on the WD's magnetic field strength ($B$) \citep{1995cvs..book.....W}. The strong magnetic field strength of a WD in polars ($B\approx10\text{--}250$\,MG) can prevent the formation of an accretion disk, and matter falls directly onto the surface of a WD.  In the case of IPs, a moderate magnetic field of a WD ($B\lesssim1$\,MG) can disrupt the formation of only the inner part of the accretion disk. In non-magnetic CVs, the disk meets the WD surface, and a boundary layer is formed. Some CVs undergo stable or unstable hydrogen or helium burning on the surface of the WD \citep[][]{2007ApJ...663.1269N,2013ApJ...777..136W}, resulting in observed phenomena such as nova outbursts, recurrent novae, and, in the single-degenerate scenario, Type Ia supernovae \citep{1973ApJ...186.1007W,1982ApJ...253..798N}. Being one of the most common classes of X-ray sources in our Galaxy, CVs are thought to significantly contribute to the Galactic ridge X-ray emission \citep{2006A&A...452..169R}. Recently, magnetic CVs have been found to make up nearly a third of all CVs \citep{2020pala, 2025rodriguez_erosita}, and exotic systems featuring rapidly evolving WDs \citep[e.g.][]{2025rodriguez_gaia22}, radio and X-ray pulsations \citep[e.g.][]{2016marsh, 2023pelisoli}, may shine a light on their complex evolution \citep[e.g.][]{2021schreiber}.

The evolution of CVs is governed by angular momentum loss, primarily through gravitational wave radiation and magnetic braking, which drives orbital shrinkage and leads to shorter orbital periods \citep{1967AcA....17..287P}.  For hydrogen-dominated CVs accreting from main-sequence donors, the observed minimum orbital period is $\rm \approx82$ minutes \citep{2009MNRAS.397.2170G}, while theoretical models predict a shorter minimum of $\rm \approx76$ minutes \citep{2006MNRAS.373..484K}. The orbital period of a majority of CVs varies from $\approx$ 80 minutes up to 10 hours. In rare cases, CVs have orbital periods below the canonical period minimum. The donor in such a CV must be denser than a normal low-mass star at the hydrogen burning limit. This can occur if the donor has evolved significantly and formed a helium-rich core before overflowing its Roche lobe \citep{1985SvAL...11...52T}. The evolved CV donor is hotter and smaller compared to normal CV donors \citep[e.g.,][]{2021MNRAS.508.4106E,2021MNRAS.505.2051E}. Evolved CVs may evolve towards shorter orbital periods and become AM Canum Venaticorum (AM CVn) systems \citep[e.g.,][]{2003MNRAS.340.1214P,2023MNRAS.519.2567S,2023MNRAS.520.3187S,2023A&A...678A..34B}. AM CVns are ultra-compact binaries and have orbital periods in the 5.4--67.8 minutes range \citep[see e.g.,][]{2010PASP..122.1133S,2018A&A...620A.141R}. Their optical spectra are dominated by helium emission lines, which makes them easily distinguishable from other CVs. Around a hundred of the AM CVn stars are currently known \citep{2018A&A...620A.141R,2025arXiv250510535G}. Among the known AM CVn systems, only two have been suggested to host moderately magnetic WDs \citep[$10^4 - 10^5$ G;][]{2024maccarone}.

Alternatively, CVs can reach periods below the canonical period minimum if they form from low-metallicty ("Population II") stars, which are more compact than solar-metallicity stars of the same mass \citep{1990ApJ...356..623H,1997A&A...320..136S}. Due to lower metallicity, Population II stars might have a period minimum of 51--67 minutes \citep{1997A&A...320..136S}, which is shorter than for Population I CVs.

The Gaia Photometric Science Alerts Team reported Gaia19bxc\footnote{ Celestial coordinates of Gaia19bxc are RA=$17^{h} 31^{m} 58^{s}.5$ and DEC=$+27\degr 09^{'} 36^{''}.2$ (Gaia DR3, \cite{2023A&A...674A...1G}) } as a transient object on 23 May 2019 \citep{2019TNSTR.838....1D,2021A&A...652A..76H}. \citet{2022arXiv220404603K} showed that Gaia19bxc exhibits high- and low-state variability, along with a 64.42-minute period in the Zwicky Transient Facility (ZTF) data. This behavior led \citet{2022arXiv220404603K} to suggest that Gaia19bxc was likely a magnetic CV with a strong magnetic field, known as a polar. In this Letter, we show that Gaia19bxc has both a photometric and \textit{spectroscopic} 64.42-minute period, strongly suggesting that this is the orbital period of the system. Phase-resolved spectroscopy reveals that this is a H-rich (not He-rich) system as well as evidence cyclotron beaming. This suggests that Gaia19bxc is a polar, and would therefore have a magnetic field strength $B\gtrsim10$ MG. This is the first such system; of the known H-rich CVs\footnote{\label{ref:evolvedCVs} CVs with orbital periods below the canonical minimum include V485 Cen \citep{1993A&A...267L..55A}, EI Psc \citep[or 1RXS J232953][]{2002ApJ...567L..49T}, CSS 100603 \citep{2012MNRAS.425.2548B}, CSS 120422 \citep{2013MNRAS.431..372C}, V418 Ser \citep{2015ApJ...815..131K},  KSP-OT-201701a \citep{2022ApJ...925L..22L}, ZTF J1813+4251 \citep{2022burdge}, and KSP-OT-201712a \citep{2024ApJ...964..186L}. The extended list can be found in \cite{2025arXiv250510535G}. } with orbital periods $P_{\rm orb} \lesssim $ 78 minutes \citep[e.g.,][]{2020MNRAS.496.1243G}, none have been found to host a strongly magnetic WD. We present optical observations of Gaia19bxc, including high-speed photometry with Hale Telescope/CHIMERA and phase-resolved spectroscopy using Keck I/LRIS (see Section \ref{sec:data}) We discuss the evolutionary stage of Gaia19bxc in Section~\ref{sec:discussion}.

\section{Data and Analysis}
\label{sec:data}

\subsection{Archival ZTF data and High-Speed Optical Photometry with CHIMERA}

\begin{figure*}
    \centering
    \includegraphics[width=0.95\textwidth]{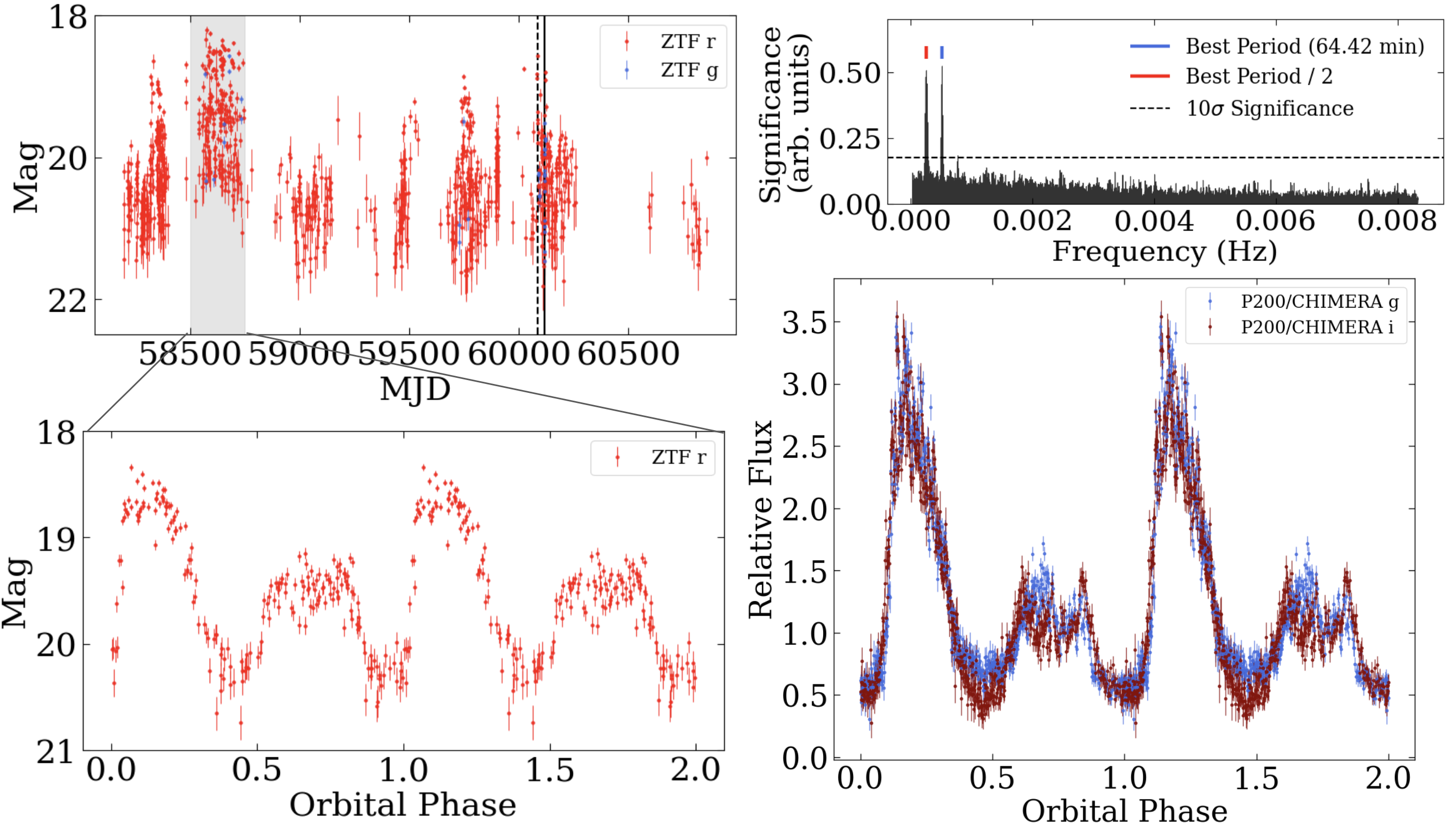}
    \caption{Archival ZTF photometry data and follow-up high-speed photometry of Gaia19bxc with CHIMERA. Upper left: The six-year-long ZTF light curve shows high and low state changes typically seen in magnetic CVs. The dashed line indicates when CHIMERA photometry was acquired and the dashed lines indicate Keck I/LRIS spectroscopy. Lower left: A phase-folded ZTF light curve. A clear period is identified (64.42 minutes) while the system is in a high state. Upper right: The Lomb-Scargle periodogram of the ZTF data shows no significant periods (e.g., WD spin period) aside from 64.42 minutes. Lower right: A phase-folded CHIMERA light curve in $g,i$ filters. Multi-band high-speed (10-sec) photometry confirms the double-peaked nature of the light curve from cyclotron emission and reveals no eclipses of the accreting WD.}
    \label{fig:LC}
\end{figure*}

The top left panel of Figure \ref{fig:LC} shows the long-term optical ZTF light curves (r- and g-filters) of Gaia19bxc. We selected the part of the long-term light curve at high state between the time of  MJD = (58500, 58750) days to search for the period. We used the Lomb-Scargle periodogram to search for the periodic signal ranging from 5 minutes to 1 day. We detect only two significant peaks on the Lomb-Scargle periodogram corresponding to the best period of $64.420\pm0.006$ minutes and its harmonics (half of the best-fit period) (see Figure \ref{fig:LC}, top right panel). The error of the photometric period is estimated from using a Monte Carlo analysis, attempting to recover a significant repeating signal assuming different periods. This period is consistent with one reported by \citet{2022arXiv220404603K}. We found no significant periods (e.g. WD spin period) aside from the best-fit period 64.42 minutes in the ZTF light curve. We also searched for the possible period in different parts of the long-term light curve, but found no period besides the 64.42 minutes.  The bottom left panel of Figure \ref{fig:LC} shows the phase-folded r-filter ZTF light curve with the period of 64.42 minutes. For obtaining orbital phases, $T_{\rm ref}$ in this paper is arbitrarily set to 60301.001 (MJD). The light curve shows two peaks, corresponding to brightness change of 2.5 mag at phase $\rm \phi=0.25$ and 1.5 mag at phase $\rm \phi=0.75$.

We obtained high-speed photometry of Gaia19bxc on 25 May 2023 using Caltech High-speed Multi-colour CAMERA \citep[CHIMERA;][]{chimera} on the Hale 200-inch telescope. Both data in Sloan i and g filters were acquired at a 10-second cadence simultaneously, covering about 120 minutes (about 1.5 times of the orbital period of Gaia19bxc).  We calibrated CHIMERA data using standard techniques. In Appendix~\ref{app:chimera}, we present the full high-speed photometry data. The bottom right panel of Figure \ref{fig:LC} shows the CHIMERA g and i filter high-speed photometry light curves of Gaia19bxc folded over the 64.42-minute orbital period. This confirms that the 64.42-minute period is the orbital period of Gaia19bxc. Similarly to archival ZTF data, CHIMERA’s light curves show two peaks at approximately phases of $\rm \phi=0.25$ and $\rm \phi=0.75$. Such a behavior of the light curve is typical for polars, showing variability within a single orbital period due to cyclotron beaming. According to the two-component cyclotron model from \cite{2008ApJ...672..531C}, two peaks in a single orbital period of Gaia19bxc can be due to the viewing angles of the system or a complex magnetic field structure \citep[see also][]{2025A&A...696A.242V}. A detailed light-curve modeling of the  Gaia19bxc is beyond the scope of this Letter.

\subsection{Phase-Resolved Optical Spectroscopy and Doppler Tomography}

\begin{figure*}
    \centering
    \includegraphics[width=0.98\textwidth]{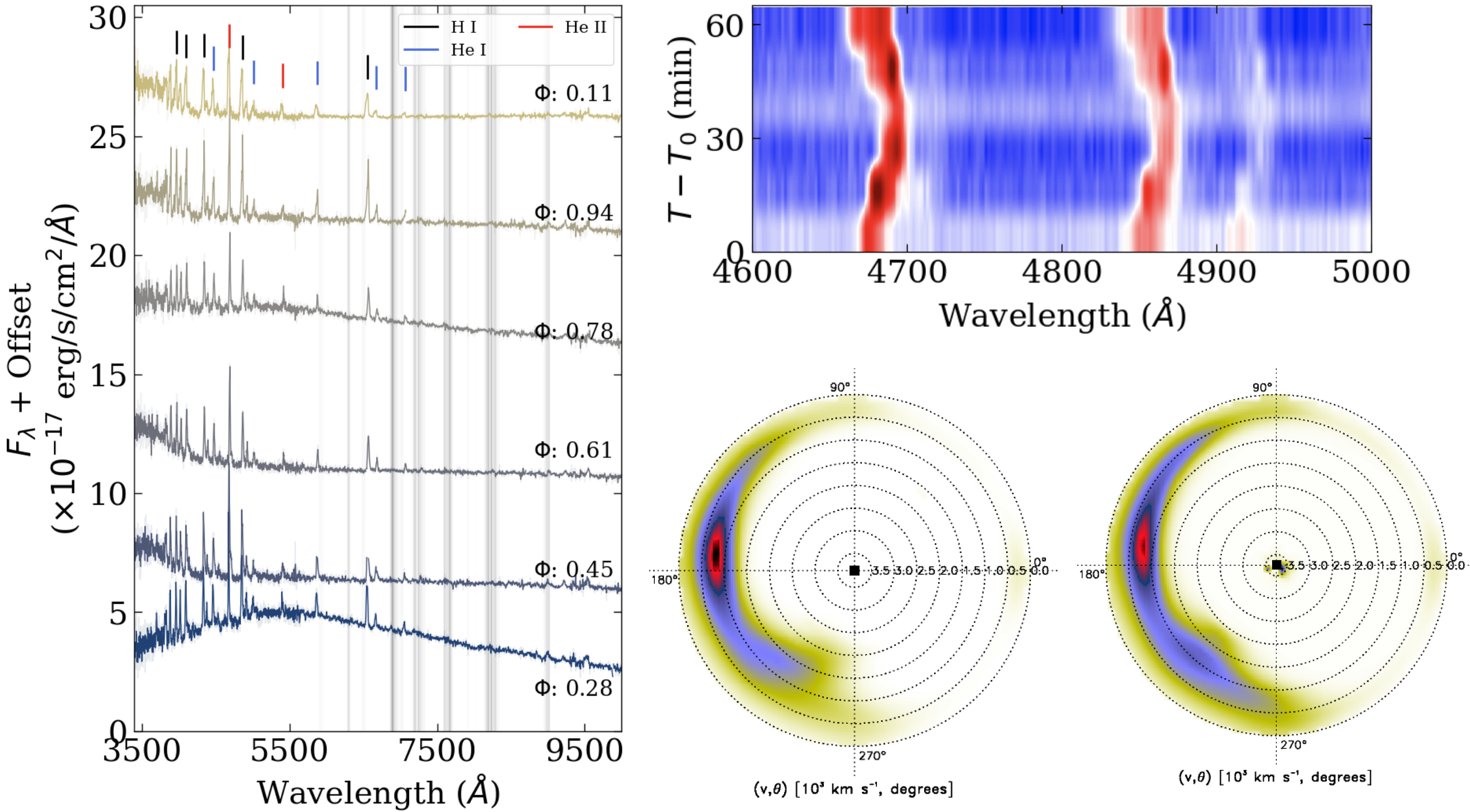}
    \caption{ Left: Keck I/LRIS phase-resolved spectroscopy of Gaia19bxc. The corresponding orbital phases for each spectrum are indicated by the text. Spectra exhibit prominent hydrogen Balmer emission lines along with high-excitation helium lines (He I and He II). A broad cyclotron hump is observed at phase of  $\rm \phi=0.28$. A small discontinuity in the spectra is caused by cosmic rays in the data. Upper right: Continuum-normalized trailed spectra of the strong He~II~4686\AA\ and H$\beta$ emission lines, and weak He~I~4921\AA. Lower right: Inverse Doppler tomograms of the same lines. The tomograms show no evidence of a disk structure but instead reveal a stream-like feature.}
    \label{fig:spectra}
\end{figure*}

\begin{table}
	\centering
	\caption{Equivalent widths (EWs) and line ratios of selected emission lines in the phase-averaged spectrum of Gaia19bxc.
	}
	\label{tab:EW_lines}
	\begin{tabular}{lc}
		\hline
		Line (\AA) &   {\parbox{2cm}{\vspace{5pt}\centering  --EW (\AA)}} \\
		\hline
\textit{Emission lines} &\\
H$\alpha$\ 6562.8 & $51.1\pm6.4$\\
H$\beta$\ 4861.3 & $45.9\pm2.1$\\
H$\gamma$\ 4340.5 & $35.4\pm3.2$\\
H$\delta$\ 4101.7 & $32.5\pm2.0$\\
H$\epsilon$\ 3970.1 & $22.6\pm2.3$\\
He I\ 4471.4  & $32.6\pm4.1$\\
He I\ 5015.7 & $30.5\pm2.1$\\
He I\ 5876.5  & $41.2\pm6.3$\\
He I\ 6678.2 & $31.3\pm2.9$\\
He I\ 7065.2  & $30.9\pm2.3$\\
He II\ 4685.7  & $46.8\pm3.3$\\
\hline
He I\ 5876.5/H$\alpha$  &  $0.81\pm 0.16$ \\
He II\ 4685.7/H$\beta$  & $1.02 \pm0.09$ \\
\hline
	\end{tabular}
\end{table}

We obtained a phase-resolved spectroscopy of Gaia19bxc over the photometric period on 23 June 2023 with Keck I telescope using the Low-Resolution Imaging Spectrometer \citep[LRIS;][]{1995PASP..107..375O}. The 600/4000 grism ($\rm 2\times2$ binning) was used on the blue side, and the 400/8500 grism on the red side ($\rm 2\times1$ binning). A $\rm 1.0 \arcsec$ slit was used and the seeing during the observation was approximately  $\rm 1.2 \arcsec$, leading to possible slit losses. Light cirrus clouds may have additionally led to a reduced throughput. We reduced the Keck I/LRIS data using standard techniques. The data were wavelength calibrated with internal lamps, flat fielded corrected, and cleaned for cosmic rays with \texttt{lpipe}, which is a pipeline optimized for LRIS imaging and long slit spectroscopy \citep{2019PASP..131h4503P}.

The left panel of Figure \ref{fig:spectra} shows phase-resolved spectra of Gaia19bxc. Orbital phases on the left panel are obtained from $T_{\rm ref}$ as specified earlier, and $T_0$ in the right panel of Figure \ref{fig:spectra} represents the start of the observation. These spectra show prominent hydrogen Balmer series along with high excitation helium (He~I and He~II) emission lines. We estimated the spectroscopic period of Gaia19bxc from radial velocity measurements of selected prominent hydrogen and helium emission lines (see Appendix~\ref{app:rv}). The spectroscopic period of $63.0\pm3.9$ minutes is consistent with the photometric period of $64.420\pm0.006$ minutes measured from CHIMERA light curves. We interpret this as a clear sign that the two are consistent with the orbital period of the system, confirming its nature well below the CV period minimum. 

We clearly observe a broad, single cyclotron hump covering effectively all optical wavelengths in the spectrum at phase of $\rm \phi=0.28$, indicating the presence of a magnetic WD in the system. Based on similar characteristics seen in other polars, this implies that Gaia19bxc  would have a magnetic field strength $B\gtrsim10$ MG, though the non-detection of individual cyclotron humps prevents a more precise constraint on the magnetic field. The emission lines in the  phase-resolved spectra are single-peaked, suggesting that the accretion of donor matter occurs via a stream directly onto the poles of the WD. We do not detect metal absorption lines from the donor's atmosphere or emission lines from the irradiated face of the donor. The absence of the donor's signature in the optical spectra indicates that it is a cold, late-type star. Table \ref{tab:EW_lines} shows the equivalent widths of the selected emission lines computed from the phase-average Keck~I/LRIS spectrum. We detect strong high-excitation He~II line, yielding a line ratio of  $\rm He\ II / H_{\beta}=1.02 \pm0.09$. Such a line ratio is commonly seen in magnetic CVs, specifically in polars \citep[e.g.][]{1992silber}.

The top right panel of Figure \ref{fig:spectra} shows the trailed, continuum-normalized Keck~I/LRIS spectra for the He~II~4686\AA\ and H$\beta$ emission lines. Over the entire 64.42-minute orbital period, the radial velocity shifts of these lines are about 500 km/s, typical for polars. We also present Doppler tomograms for the He~II~4686\AA\ and H$\beta$ lines, which map line shifts and strengths into velocity as a function of orbital phase \citep[e.g.][]{1988marsh}. We used the \texttt{doptomog}\footnote{\url{https://www.saao.ac.za/~ejk/doptomog/main.html}} code developed by \cite{2015kotze} to construct Doppler tomograms shown here. The ``inverse''\footnote{This refers to the manner in which velocities are shown in polar coordinates: ``inverse'' Doppler tomograms show higher velocities inside, lower velocities outside.} Doppler tomograms for these emission lines reveal no disk structure. Instead, a stream-like feature is observed, resembling those commonly seen in polars (see Figure \ref{fig:spectra}, bottom right panel).

\section{Discussion and Conclusion}
\label{sec:discussion}

\begin{figure}
    \centering
    \includegraphics[width=0.45\textwidth]{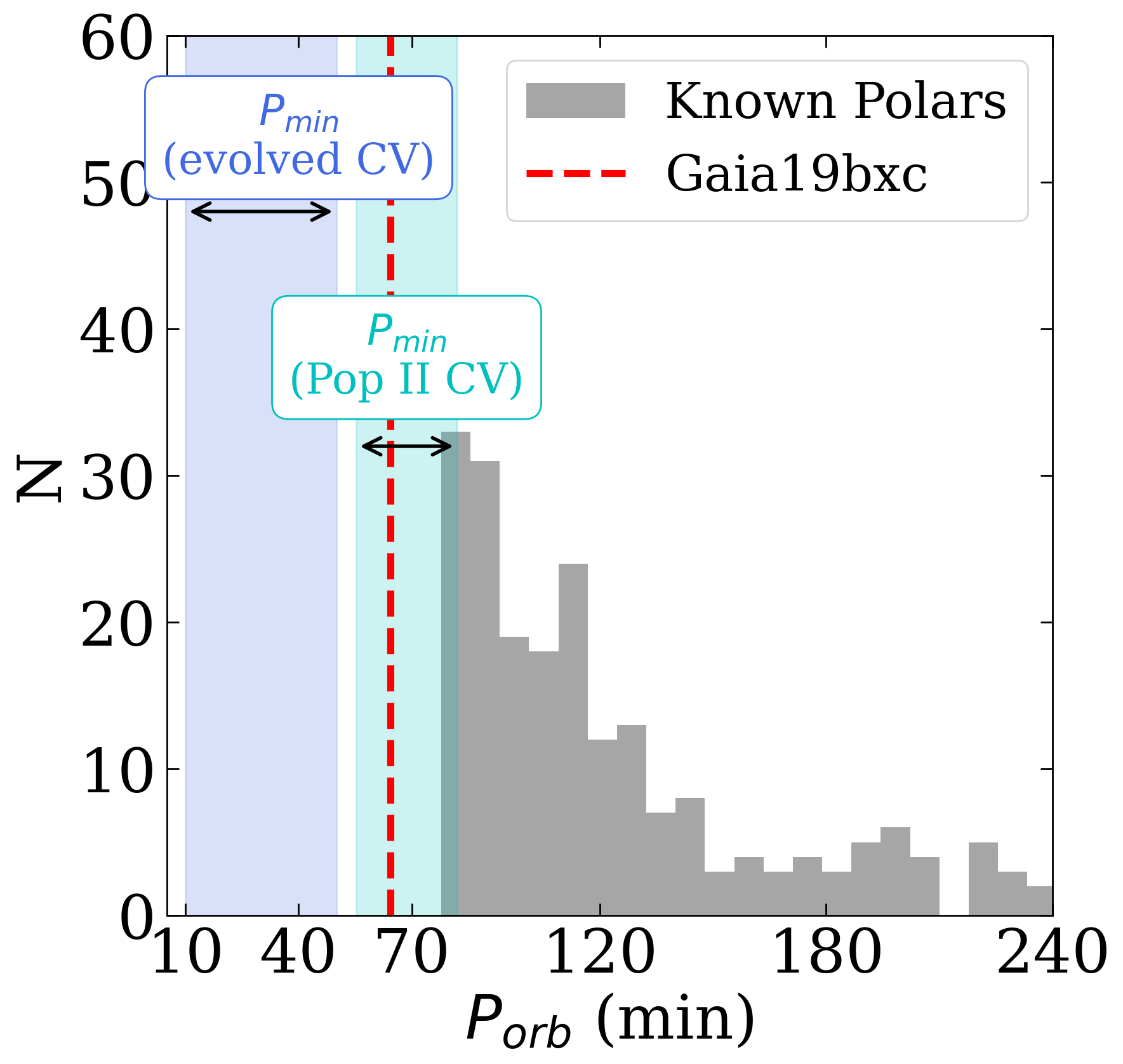}
    \caption{The period distribution of known polars from \cite{2003A&A...404..301R} (gray). The orbital period of Gaia19bxc, with its 64.42-minute period is marked by the red vertical line, showing that Gaia19bxc is the most compact polar known to date. The cyan region indicates the predicted $P_{\rm min}$ range ($\rm 51-78$ minutes) for Population II (metal-poor) CVs \citep[see Table 2 of ][]{1997A&A...320..136S}. The blue region represents the predicted $P_{\rm min}$ range ($\rm 10-50$ minutes) for evolved CVs from the MESA models of \cite{2021MNRAS.508.4106E}. Since both evolutionary paths lead to period minima below the orbital period of Gaia19bxc, a clear detection and characterization of the donor of Gaia19bxc would be crucial to distinguish between these two scenarios.}
    \label{fig:per_distr}
\end{figure}

High-speed photometry and phase-resolved spectroscopy indicate that Gaia19bxc is a polar with an orbital period of 64.42 minutes. Such an orbital period is well below the canonical period minimum  ($P_{\rm min}$) for CVs ($\rm \approx76$ minutes, \citealt{2006MNRAS.373..484K}; $\rm \approx82$ minutes, \citealt{2009MNRAS.397.2170G}), suggesting that Gaia19bxc is in a unique evolutionary stage. In Figure~\ref{fig:per_distr}, we show that the orbital period of Gaia19bxc, while not above the  $P_{\rm min}$ for typical CVs, is above that of either Population II CVs or evolved CVs. Below, we discuss three possibilities for Gaia19bxc being: {\it (a)} an AM CVn, {\it (b)} an evolved CV, or {\it (c)} a Population II CV.

{\it (a) AM CVn system:} Gaia19bxc displays a distinctive spectroscopic signature, with optical spectra revealing both helium and hydrogen emission lines of comparable strength. Due to its ultra-compact nature, one may have classified it as an AM CVn had spectroscopy not been acquired; here, we showed that this scenario is incorrect.

{\it (b) Evolved CV:} CVs can reach short orbital periods (below the canonical period minimum) if their donors have exhausted a significant fraction of the hydrogen in cores before filling the Roche lobe \citep{1985SvAL...11...52T}. The donor of such CVs is significantly evolved, smaller, and hotter, compared to normal CVs. Only around a dozen of CVs with orbital periods below the canonical minimum are known, likely harboring evolved donors\footref{ref:evolvedCVs}. These systems exhibit a relatively high helium-to-hydrogen line ratio, $\rm He\ I\ 5876.5/H\alpha \lesssim 1$, comparable to Gaia19bxc, which has $\rm He\ I\ 5876.5/H\alpha=0.81$ (see Table~\ref{tab:EW_lines}). According to the {\it MESA} binary models for CVs with evolved donors \citep{2021MNRAS.508.4106E,2021MNRAS.505.2051E}, the donor in Gaia19bxc should be hot, with a temperature $\rm \gtrsim 4000$ K (see Figure~\ref{fig:app_SED} in Appendix~\ref{app:SED}). Gaia19bxc may evolve toward a shorter orbital period via the "evolved CV" channel, potentially becoming a magnetic AM CVn system. However, we find no evidence of a hot donor in the phase-resolved spectra of Gaia19bxc, which suggests that the donor is cooler than expected for an evolved CVs (see Appendix~\ref{app:SED} for more details). 

{\it (c) Population II (metal-poor) CV:} Population II CVs represent old CVs with low-metallicity donors, formed during the early phases of Galactic evolution \citep{1990ApJ...356..623H,1997A&A...320..136S}. These CVs are predominantly located in the Galactic halo, which is metal-poor. Due to the low metallicity of their donors, Population II CVs exhibit shorter evolutionary timescales, and their period minimum varies from 51 to 59 minutes for metallicities of  $Z = 10^{-5}$ to $10^{-3}$ \citep[see Table 2,][]{1997A&A...320..136S}. Being a member of the Galactic halo, Population II CVs exhibit high space velocities and proper motions. We compare the kinematic properties of Gaia19bxc to those of a known Population II CV, SDSS J150722.30+523039.8 \citep[SDSS J15072,][]{2008PASP..120..510P,2011MNRAS.414L..85U}. According to the Gaia DR3 catalog, SDSS J15072 is located at Galactic coordinates $\rm (l,b)=(87.38\degr, 54.19\degr)$, and it shows a high proper motion of $\rm 154.84\ mas/yr$ and a transverse velocity of approximately  $\rm \approx 156\ km/s$. In comparison, Gaia19bxc is located at Galactic coordinates $\rm (l,b)=(50.72\degr, 28.50\degr)$, and it exhibits a high proper motion of $\rm 9.62\ mas/yr$ with a transverse velocity of approximately $\rm \approx 91\  km/s$ (assuming a distance of 2 kpc; see Appendix~\ref{app:SED}). These kinematic properties of Gaia19bxc are comparable to those of Population II CVs.  With an orbital period of 64.42 minutes, Gaia19bxc might be a pre-bounce system or a period-bouncer\footnote{Period bouncers are CVs whose orbital periods have passed the period minimum during their evolution and are now increasing toward longer periods.} of Population II CVs.

In summary, Gaia19bxc, with an orbital period of 64.42 minutes, is the first polar below the canonical period minimum of CVs. We found no evidence of a donor in the phase-resolved spectra, such as absorption lines originating from the donor's atmosphere or emission lines arising from the irradiated surface of the donor. This suggests that the donor of Gaia19bxc is cold. If Gaia19bxc were an evolved CV, the donor should be a hot, and we would expect to observe its signatures in optical spectra. Based on the discussion above, we favor the interpretation that Gaia19bxc is a Population II CV rather than an evolved CV. Our understanding of such systems is crucial since Gaia19bxc is at the faintest end ($\approx$20--21 mag) of what current photometric surveys such as ZTF can detect. In the coming years, the Rubin Observatory Legacy Survey of Space and Time \citep[LSST; ][]{2019rubin}, reaching 23--24 mag, is bound to discover dozens more of these intrinsically faint systems, therefore increasing our understanding of magnetic field generation and evolution in close binary stars.

\begin{acknowledgements}

Based on observations obtained with the Samuel Oschin Telescope 48-inch and the 60-inch Telescope at the Palomar Observatory as part of the Zwicky Transient Facility project. ZTF is supported by the National Science Foundation under Grants No. AST-1440341 and AST-2034437 and a collaboration including current partners Caltech, IPAC, the Weizmann Institute of Science, the Oskar Klein Center at Stockholm University, the University of Maryland, Deutsches Elektronen-Synchrotron and Humboldt University, the TANGO Consortium of Taiwan, the University of Wisconsin at Milwaukee, Trinity College Dublin, Lawrence Livermore National Laboratories, IN2P3, University of Warwick, Ruhr University Bochum, Northwestern University and former partners the University of Washington, Los Alamos National Laboratories, and Lawrence Berkeley National Laboratories. Operations are conducted by COO, IPAC, and UW. This work has made use of data from the European Space Agency (ESA) mission Gaia (\url{https://www.cosmos.esa.int/gaia}), processed by the Gaia Data Processing and Analysis Consortium (DPAC, \url{https://www.cosmos.esa.int/web/gaia/dpac/consortium}). Funding for the DPAC has been provided by national institutions, in particular the institutions participating in the Gaia Multilateral Agreement. Some of the data presented herein were obtained at Keck Observatory, which is a private 501(c)3 non-profit organization operated as a scientific partnership among the California Institute of Technology, the University of California, and the National Aeronautics and Space Administration. The Observatory was made possible by the generous financial support of the W. M. Keck Foundation. We wish to recognize and acknowledge the very significant cultural role and reverence that the summit of Maunakea has always had within the Native Hawaiian community. We are most fortunate to have the opportunity to conduct observations from this mountain. We are grateful to the staffs of the Palomar and Keck Observatories for their work in helping us carry out our observations. 

IG acknowledges support from Kazan Federal University. ACR acknowledges support from the National Science Foundation via an NSF Graduate Research Fellowship. We thank the anonymous referee for useful comments and suggestions, which contributed to the improvement of this manuscript.

\end{acknowledgements}

\appendix

\section{Full CHIMERA high speed photometry light curves}
\label{app:chimera}

\begin{figure}[]
    \centering
    \includegraphics[scale=0.3]{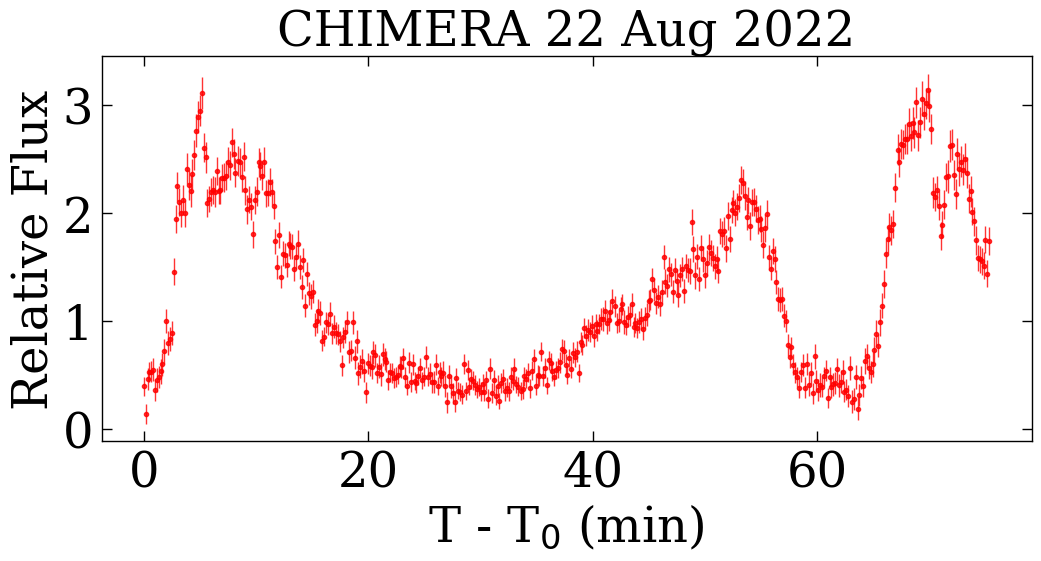}
    \includegraphics[scale=0.3]{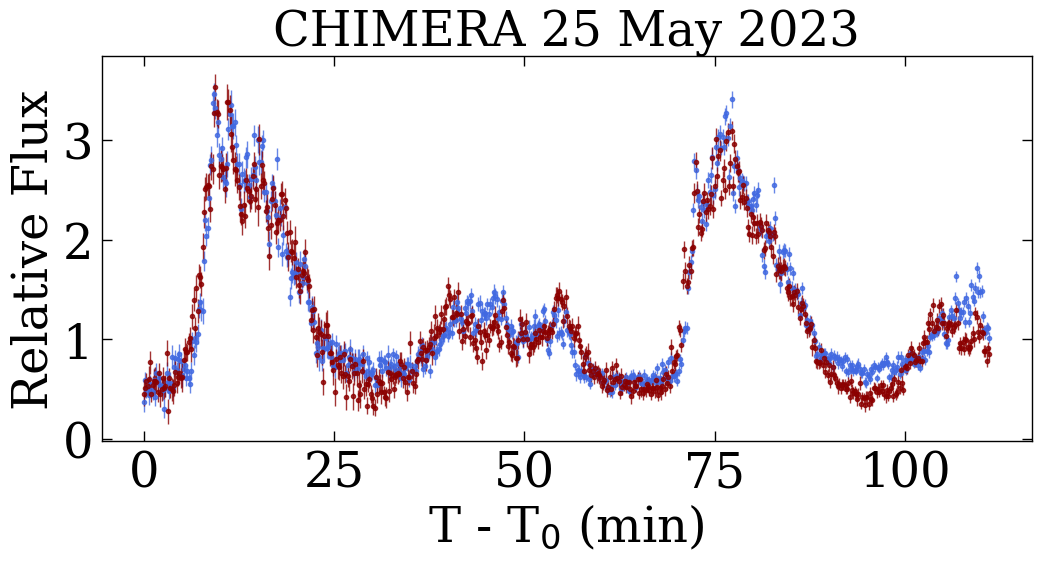}
    \caption{Full CHIMERA high-speed photometry light curves in $g$, $i$, $r$ filters covering the entire orbital period of Gaia19bxc: 22 August 2022 ($r$ filter, top panel), and 25 May 2023 ($g$, $i$ filters, bottom panel; blue and dark red, respectively). The orbital period of 64.42 minutes for Gaia19bxc is confirmed, which is below the period minimum for CVs.}
    \label{fig:app_chimera}
\end{figure}

We calibrated CHIMERA data using standard techniques in the PyCHIMERA pipeline (bias-subtraction, flat-field correction). Aperture photometry was performed with the ULTRACAM pipeline \citep{2007ultracam}. Relative fluxes were computed by dividing the target fluxes by those of a standard star. Figure \ref{fig:app_chimera} shows continuous CHIMERA light curves in $g$, $i$, $r$ filters covering the entire orbital period of Gaia19bxc. Approximately 90 minutes of observations with CHIMERA were obtained on 22 August 2022 ($r$ filter) and 120 minutes on 25 May 2023 ($g$, $i$ filters). The light curves were shifted to align the same phase at the starting point. The orbital period of 64.42 minutes for Gaia19bxc is confirmed with these data. The light curves in the $g$, $i$ filters display a double-peaked structure at phases $\phi \approx 0.25$ and $\phi \approx 0.75$ (see Figure \ref{fig:LC}). The $r$-filter light curve shows a plateau at $\phi \approx 0.75$ (or  $(T - T_0) \approx 50$ minutes in Figure~\ref{fig:app_chimera}) followed by a continued rise in brightness up to phase of $\phi \approx 0.8$ (or $(T - T_0) \approx55$ minutes in Figure \ref{fig:app_chimera}). Such differences in the light curve shapes suggest contributions from cyclotron emission.

\section{ Radial velocity and spectroscopic period determination}
\label{app:rv}

\begin{figure}[]
    \centering
    \includegraphics[width=0.329\textwidth]{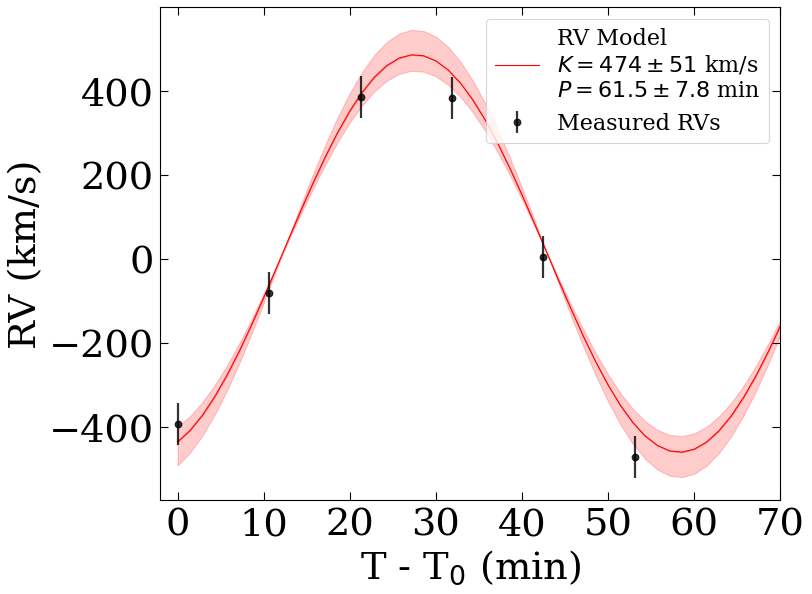}
    \includegraphics[width=0.33\textwidth]{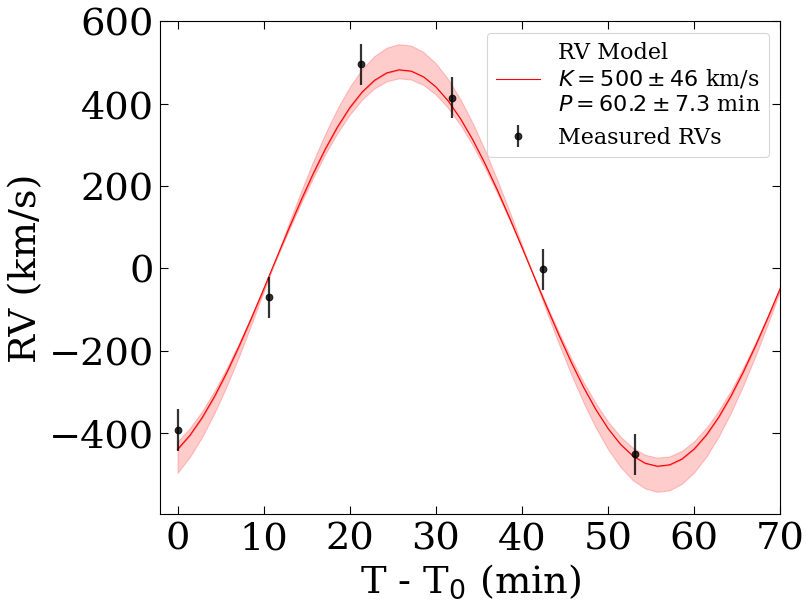}
    \includegraphics[width=0.33\textwidth]{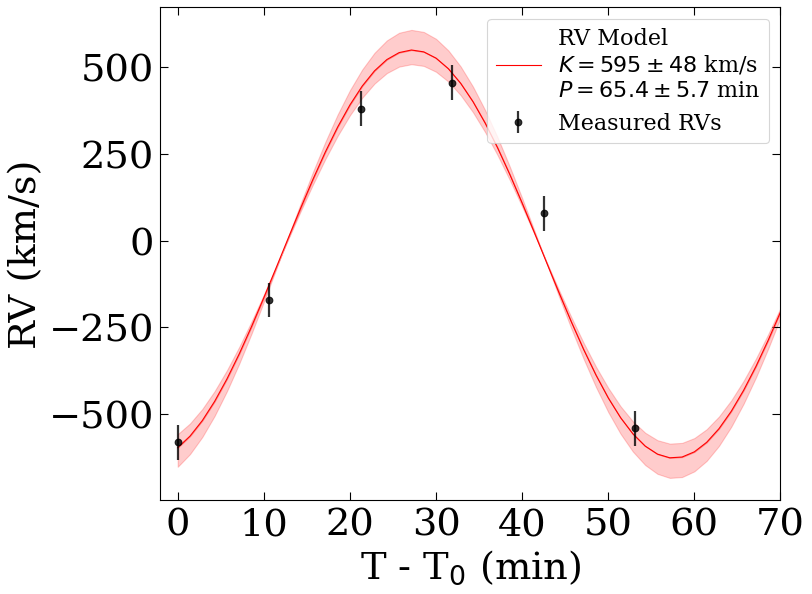}
    \caption{ RV measurements (black points) of prominent hydrogen (H$\beta$, H$\gamma$) and helium (He~II~4686\AA) emission lines, from left to right. Solid red lines represent the RV model constructed from the median parameters of the MCMC parameter exploration. Shaded red regions indicate models constructed from the 1$\sigma$ uncertainties. The mean spectroscopic period derived from their weighted average is $63.0\pm3.9$ minutes, which is consistent with the photometric period (64.42 minutes). We interpret this as the orbital period, which confirms the polar nature of the source.}
    \label{fig:app_rv}
\end{figure}

We used prominent hydrogen (H$\beta$, H$\gamma$) and helium (He~II~4686\AA) emission lines to estimate the period of Gaia19bxc from spectroscopy alone. We fitted single-peaked Gaussian profiles to those emission lines in each phase-resolved spectrum and derived radial velocity (RV) measurements and errors from the central wavelength of the Gaussians and error of the mean, respectively. Assuming a circular, Keplerian orbit, we used a sinusoidal function of the form:

\begin{equation}
\mathrm{RV} =  K \sin\left( \frac{2\pi (T - T_0)}{P} \right) + \gamma,
\end{equation}

where $K$ is the semi-amplitude of the emission line, $\gamma$ is the systemic velocity, and $P$ is the spectroscopic period. We used a Markov chain Monte Carlo (MCMC) Bayesian parameter exploration \citep{1970Bimka..57...97H} to generate posterior distributions of all parameters and their errors. The MCMC was performed using the {\it emcee} package \citep{2013PASP..125..306F} with 15000 runs and twelve walkers, with the first half used as the burn-in period. Uniform priors were used for all parameters, where $K$ was allowed to range between 0 and 800 km/s, $\gamma$ between -500 and 500 km/s, and $P$ between 40 and 100 minutes. Figure \ref{fig:app_rv} shows the RV curves for all emission lines. By taking the weighted mean, we determined the spectroscopic period of Gaia19bxc to be $63.0\pm3.9$ minutes. This spectroscopic period is consistent with the photometric orbital period derived from CHIMERA data, the latter of which is measured to a much greater precision.

\section{SED constraints on donor temperature: an unlikely evolved donor scenario}
\label{app:SED}

\begin{figure}
    \centering
    \includegraphics[width=\textwidth]{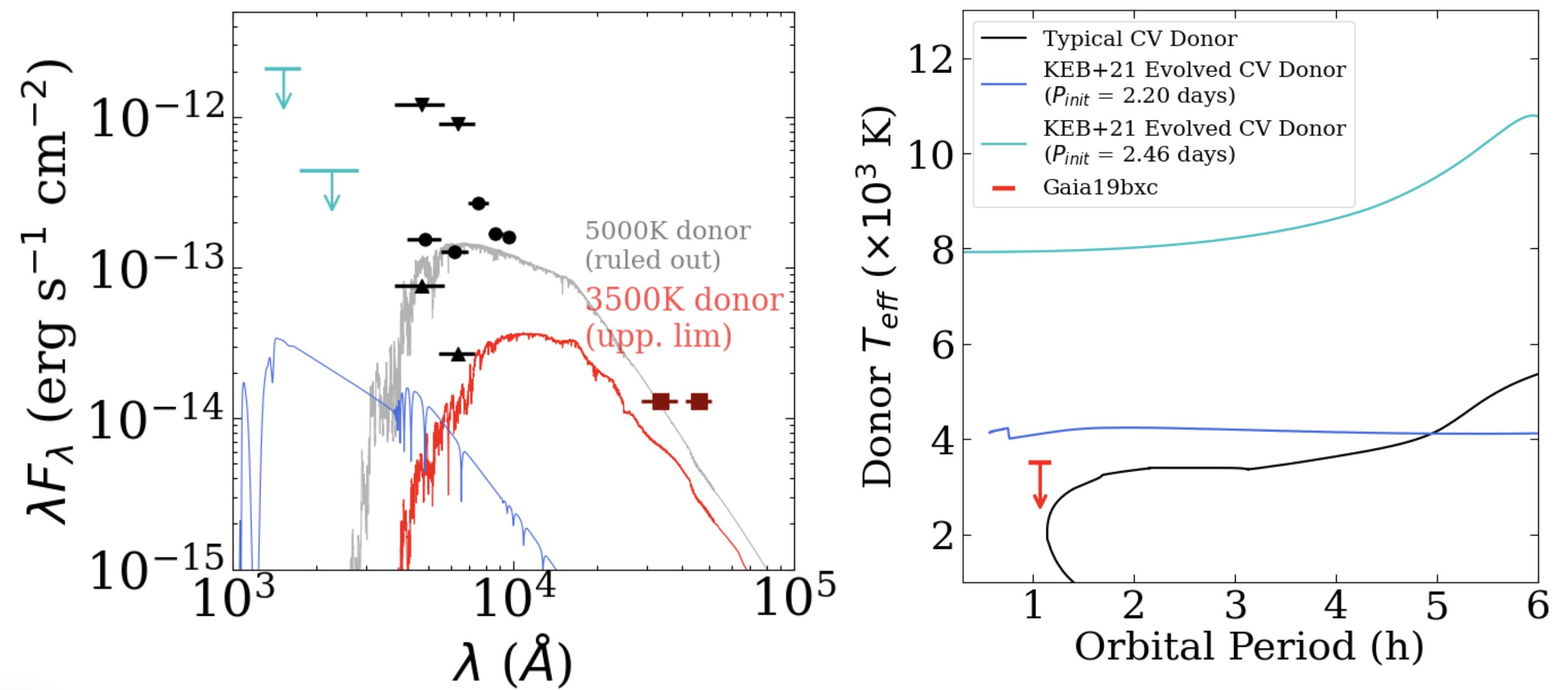}
    \caption{Left: Observed SED of Gaia19bxc. Photometric data from GALEX (UV, cyan), PanSTARRS PS1 (optical, black circles), ZTF (optical, black squares), and CatWISE (mid-IR, dark red) are shown. The ZTF data at the light curve minimum (upward triangles) and maximum (downward triangles) are indicated in the same bandpass but with different fluxes. The blue curve represents a WD model atmosphere with $T_{\rm eff} \approx 15 000$ K. Adopting a fiducial distance of 2 kpc, the donor has an upper effective temperature limit of $T_{\rm eff} \lesssim 3500$ K, based on the ZTF point at light curve minimum. A hotter donor with an effective temperature of $T_{\rm eff} \approx 5000$ K would overestimate the ZTF $r$-filter flux (at minimum) and should be detectable in phase-resolved spectra; however, such a donor is ruled out based on current observations. Right: Donor effective temperature versus orbital period plane. \textit{MESA} binary models for evolved CV donors with initial periods $\rm P_{init} = 2.20$ (blue) and $\rm P_{init} = 2.46$ days (cyan) are adopted from \citet{2021MNRAS.508.4106E,2021MNRAS.505.2051E}. A model with $\rm P_{init} = 1.00$ days (black) represents typical CVs with unevolved donors. The donor effective temperature of $T_{\rm eff} \lesssim 3500$ K places Gaia19bxc below evolved CV models (red).}
    \label{fig:app_SED}
\end{figure}

We constructed the observed spectral energy distribution (SED) of Gaia19bxc based on photometric data from the Galaxy Evolution Explorer \citep[GALEX;][]{2005ApJ...619L...1M}, ZTF and CatWISE catalog \citep{2020ApJS..247...69E}. No archival near-infrared photometry, X-ray, or radio detections are available for Gaia19bxc in the literature. Gaia19bxc does not have a well-constrained distance, since it has a \textit{Gaia} DR3 parallax of $\pi = 0.57\pm 0.77$ mas.  While the geometric distance from \cite{2021AJ....161..147B} is $d=2009^{+832}_{-960}$ pc, this value should be used with caution, since the parallax error exceeds the parallax measurement itself. For further analysis we adopted the fiducial distance of 2 kpc. We adopted an extinction of $\rm A_v = 0.2$, obtained from the Bayestar19 dust map of \cite{2019green_map}.

The left panel of Figure \ref{fig:app_SED} shows the observed SED of Gaia19bxc. The photometric data on the SED are stochastically distributed, indicating that the data are dominated by cyclotron emission at different phases of Gaia19bxc. Direct SED modeling requires the consideration of a cyclotron emission model along with WD and donor atmosphere models. We constrained the donor temperature under the assumption that all emission originates from the WD and the donor. The donor star effective temperature is specified (see Figure \ref{fig:app_SED}), interpolating through BT-NextGen library of theoretical stellar atmospheres \citep{2011allard}, assuming solar metallicity and $\log g = 5.0$. The stellar atmosphere is multiplied by $(R/d)^2$ to obtain a flux as viewed from Earth. The same process is used for the WD, taking the theoretical (DA; H-rich) WD atmospheres of \citep{2010koester} and assuming $\log g = 8.0$. In the WD case, $T_{\rm WD} = 14000$ K is assumed, which is typical of WDs in CVs near the fiducial period minimum, and the typical CV WD mass of 0.8$M_\odot$ is adopted \cite{2022pala}. 

Adopting the fiducial distance of 2 kpc, the donor has an upper limit of  $T_{\rm eff}\lesssim 3500$ K. Such a donor model fits only the ZTF $r$-filter flux and underestimates the WISE and other optical photometric data (see Figure \ref{fig:app_SED}, left panel). This discrepancy can be explained by the significant contribution of cyclotron emission. Conversely, a donor model with $T_{\rm eff}\approx5000$ K fits the WISE and most of the ZTF photometric data but overestimates the ZTF $r$-filter flux at the light curve minimum. A donor with $T_{\rm eff}\approx5000$~K should be visible in optical spectra and contribute significantly to the $r$-filter flux. Since we do not observe donor features in the optical spectra, a donor with $T_{\rm eff}\approx5000$~K can be ruled out. A WISE $\rm (W_{1}-W_{2}) \approx 1.0$ color of Gaia19bxc alone also implies an effective temperature of the donor of $\rm T_{\rm eff}\sim 1500\ $~K \citep[][see Table 1]{2015A&A...574A..78S}, assuming that only the donor radiates in the infrared. 

The right panel of Figure~\ref{fig:app_SED} shows the donor effective temperature versus orbital period plane, along with \textit{MESA} binary models for CVs with evolved donors from \cite{2021MNRAS.508.4106E,2021MNRAS.505.2051E}.  Models with initial orbital periods of $\rm P_{\mathrm{init}} = 2.20$ and $\rm P_{\mathrm{init}} = 2.46$ days correspond to initially detached white dwarf plus main-sequence binaries.  For typical CVs, the initial period is $\rm P_{init} = 1.00$ day. Binaries with initial periods of $\rm P_{init} \gtrsim  2.20$ days behave differently compared to typical CVs. The donor in such systems will evolve before filling its Roche lobe. Such a CVs have an evolved donor with effective temperatures of  $\rm \gtrsim  4000$~K (for models with $\rm P_{init} \gtrsim  2.2$ days). At the end of their evolution, they become AM CVn systems via the ``evolved CV" channel. The upper limit on the donor temperature ($T_{\rm eff} \lesssim 3 500$~K) places Gaia19bxc below the evolved CV models. 

We note that we used a simplified approach to constrain the donor temperature of Gaia19bxc. A more precise analysis would require including cyclotron emission in the SED modeling. Gaia19bxc does not have a well-constrained distance, which further complicates the analysis. In our data, we do not detect any emission or absorption lines from the donor itself in the optical spectra. Future infrared spectroscopy of the donor is necessary to precisely determine its effective temperature and metallicity.

\bibliography{Gaia19bxc}{}
\bibliographystyle{aasjournal}

\end{document}